\documentclass[fleqn,12pt,twoside]{article}
\usepackage{espcrc1,epsf}

\usepackage{graphicx}
\usepackage[figuresright]{rotating}

\newcommand{\AmS}{{\protect\the\textfont2
  A\kern-.1667em\lower.5ex\hbox{M}\kern-.125emS}}
\newcommand{\msun}{M_\odot}

\def\H1{$^1$H}

\def\He4{$^4$He}
\def\Li7{$^7$Li}
\def\Be7{$^7$Be}
\def\C12{$^{12}$C}
\def\Cthirteen{$^{13}$C}
\def\N14{$^{14}$N}

\def\O16{$^{16}$O}

\def\Ne22{$^{22}$Ne}

\def\F19{$^{19}$F}
\def\Na23{$^{23}$Na}
\def\Mg24{$^{24}$Mg}
\def\Mgtf{$^{25}$Mg}

\def\Alg26{$^{26}$Al$^g$}
\def\Al26{$^{26}$Al}

\def\Si30{$^{30}$Si}
\def \la{\mathrel{\mathchoice   {\vcenter{\offinterlineskip\halign{\hfil
$\displaystyle##$\hfil\cr<\cr\sim\cr}}}
{\vcenter{\offinterlineskip\halign{\hfil$\textstyle##$\hfil\cr
<\cr\sim\cr}}}
{\vcenter{\offinterlineskip\halign{\hfil$\scriptstyle##$\hfil\cr
<\cr\sim\cr}}}
{\vcenter{\offinterlineskip\halign{\hfil$\scriptscriptstyle##$\hfil\cr
<\cr\sim\cr}}}}}
\def \ga{\mathrel{\mathchoice   {\vcenter{\offinterlineskip\halign{\hfil
$\displaystyle##$\hfil\cr>\cr\sim\cr}}}
{\vcenter{\offinterlineskip\halign{\hfil$\textstyle##$\hfil\cr
>\cr\sim\cr}}}
{\vcenter{\offinterlineskip\halign{\hfil$\scriptstyle##$\hfil\cr
>\cr\sim\cr}}}
{\vcenter{\offinterlineskip\halign{\hfil$\scriptscriptstyle##$\hfil\cr
>\cr\sim\cr}}}}}

\title{What we {\it do\/} and {\it do not\/} know about the $s$-process in AGB stars}

\author{J. C. Lattanzio\address[1]{Centre for Stellar and Planetary Astrophysics, 
        Monash University, Australia; and 
        IGPP, Lawrence Livermore National Laboratory, California, USA} and %
        M. A. Lugaro\address[3]{Institute of Astronomy, Cambridge University, UK}}
\begin{document}

\maketitle

\begin{abstract}
AGB stars are the source for the main component of the 
$s$-process. Here we discuss both the properties which 
are reasonably well known and those  which still suffer from  
substantial uncertainties. 
In the former case, we are fairly sure that the  $s$-process 
contribution from AGB stars comes from masses between about 1 and 3 $\msun$, 
and the dominant
neutron source is the $^{13}$C$(\alpha$,n)$^{16}$O reaction. In the latter
category remains the formation mechanism for the $^{13}$C-pocket.
Attempts at including rotation seem to inhibit neutron capture
reactions. Explaining the observations seems to require a spread in
the size of the $^{13}$C-pocket so some stochastic process, such as rotation, 
must be involved.
\end{abstract}

\section{Introduction to the $s$-process}

The $s$-process refers to neutron captures being $\bf s$low compared 
to subsequent beta decays, typically with $n_n \la 
10^8$~cm$^{-3}$. In this approximation, 
and assuming no branchings,
the equation governing the abundance $N_A$ of the (stable) 
isobar of mass $A$  is
\begin{equation}
{{{\rm d} N_A}\over{{\rm d} t}} = -n_n\langle \sigma v\rangle_A N_A
+ n_n\langle \sigma v\rangle_{A-1} N_{A-1}
\end{equation}
where $n_n$ is the
neutron number density and $\langle \sigma v\rangle_A$ is the
thermally averaged neutron-capture cross-section for the isobar of
mass A. It is common to write $\langle \sigma v\rangle_A $ as
$\sigma_A v_T$ where $v_T$ is the thermal velocity of the neutrons and
$\sigma_A$ is an appropriate average cross-section.
It is useful to  define the neutron exposure by
%
$\tau = \int n_n v_T \,dt\,,$
%
and thus we get
\begin{equation}
{{{\rm d} N_A}\over{{\rm d} \tau}} = -\sigma_A N_A + 
                                   \sigma_{A-1} N_{A-1}\,.
\end{equation}

In a steady-state
${\rm d}N_A/{\rm d}t = 0$ and $\sigma_A
N_A$ is constant.
The solar system distribution
does show $\sigma_A N_A$ roughly constant when
away from the isobars of magic neutron number (these
produce bottle-necks in the distribution
due to their low cross-section).

Clayton et al.  \cite{clayton:61} showed that a single neutron exposure
$\tau$ could not reproduce the solar system distribution, 
and we historically recognise three distinct components:

\begin{enumerate}
\item[1)]{\sl Weak Component:\/}~producing most of the $s$-isotopes
with $A \la 90$, from Fe to Sr;
\vspace{-0.25 cm}
\item[2)]{\sl Main Component:\/}~responsible for
the $s$-isotopes from $90 \la A \la 204$, from Sr to Pb;

\vspace{-0.25 cm}
\item[3)]{\sl Strong Component:\/}~devised primarily to produce 
$^{208}$Pb in the solar system.
\end{enumerate}

\vspace{-0.2 cm}
To reproduce the solar system distribution we add a mix of these
three components.  The weak component
is believed to come from central He burning in 
massive stars, where the neutron source is \Ne22($\alpha,$n)\Mgtf. The
main component is associated with AGB
stars, and the strong component is now associated with metal-poor AGB
stars.

\section{The $^{22}$Ne neutron Source}

There is strong
evidence that most giant stars enriched in $s$-process elements have
masses around $1-3\,\msun$. In these stars the
neutron source is \Cthirteen$(\alpha,n)$\O16, as discussed in the next 
section. But for intermediate-mass AGB stars ($M \ga 3\msun$) the neutron
source is thought to be \Ne22.

In many ways, the \Ne22 source is the simplest to activate. The H-shell
burns CNO into \N14 which can then capture two alpha-particles during 
the next thermal pulse to produce \Ne22 in the flash-driven convective 
pocket. If the peak temperature in this pocket exceeds $300$ million~K 
then neutrons are released by \Ne22$(\alpha,n)$\Mgtf. However, a few
things work against the \Ne22 source being important for the $s$-process.
Firstly there is the fact that the extent in mass of the flash-driven 
convective zone decreases as the stellar mass increases, 
from about 0.03$\msun$ (in low mass stars) to below 0.005$\msun$ for 
intermediate masses.
Thus, the number
of seeds and neutrons is small, so that not much processing can occur.
Secondly, to make the situation worse, the duration in time of the
pocket also decreases with mass, from about 300y to 20y, 
giving less time for the neutrons
to be produced. This is at least partly offset by the fact that
the temperature of the shell increases with mass, making more neutrons 
available at higher masses. Thirdly, the small convective zone of 
enriched $s$-process elements is then diluted in a large envelope 
prior to its ejection into the interstellar medium via the stellar wind. 
And finally, the shape of the IMF also works against intermediate mass stars 
being an important site for the $s$-process. Nevertheless, a quantitative 
estimate of their importance is not available at present, and would 
be useful. Further, it is possible that the \Cthirteen~source could
be active in these stars in addition to the \Ne22 source. It would 
also suffer from the extreme closeness of the H and He shells in 
intermediate-mass stars, and hence, is unlikely to be important.
But a quantitative analysis is not yet available.

\section{The $^{13}$C neutron Source}

\subsection{Basics of the $^{13}$C Source}
\begin{figure}
\begin{center}
\epsfxsize = 0.9\hsize
\centerline{\epsffile{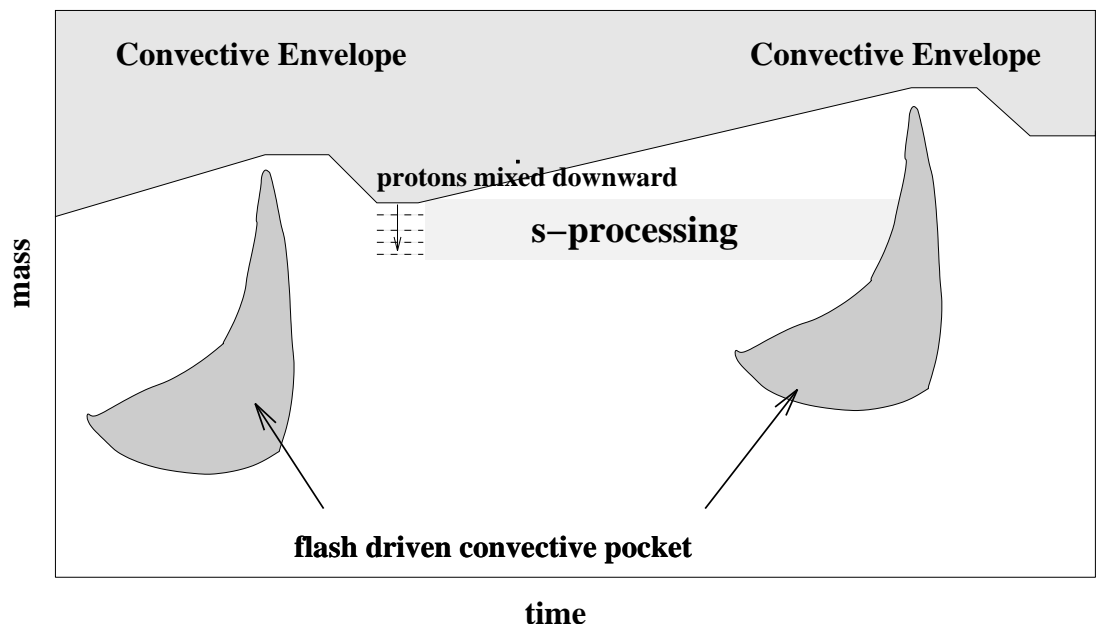}}
\smallskip
\caption{Schematic diagram showing the operation of the 
\Cthirteen\ neutron source.}
\label{c13schematic}
\end{center}
\vspace{-1cm}
\end{figure}

The basic mechanism of the \Cthirteen\ source is fairly simple, and is shown 
in Figure~\ref{c13schematic}. Some protons are mixed below the hydrogen-rich 
envelope at the time that dredge-up ceases. This region is comprised of about 
25\%~\C12~and 75\% \He4~(by mass). The protons are captured by the abundant
\C12~nuclei to form \Cthirteen ~and \N14. When the star contracts again, the
H shell reignites and the temperature in the \Cthirteen\ pocket
approaches 100 million~K where the timescale for $\alpha$-capture
decreases below the time between pulses \cite{straniero:95}. Hence, neutrons
are released within the pocket at quite low neutron densities.
These neutrons are captured {\it in situ\/} 
by Fe and heavy species to produce the $s$-process isotopes. At the next pulse 
this $s$-process rich shell is ingested by the convective pocket. In addition, 
even in low 
mass stars there might be a brief activation of the \Ne22~source at the peak 
of the pulse. This much is relatively clear. The details, however, are another matter.

\subsection{Formation of the $^{13}$C Pocket}
The largest unknown in the scenario described above is the mechanism 
which causes the protons to be mixed into the carbon-enriched region.
We will discuss four mechanism which have been suggested and explored.

\subsubsection{Semiconvection}
The first calculation which showed the activation of the \Cthirteen\  source was  
\cite{IR:82a},\cite{IR:82b}. They found that, following a pulse, the expansion of the star
caused  some recombination of fully ionized carbon atoms at the very bottom of
the hydrogen-rich envelope. This caused a dramatic change in the opacity, and 
a small semiconvective region developed at the bottom of the formally convective envelope.
This mixed some protons down and produced a thin region, a few $\times 10^{-4} \msun$, 
containing
a mass fraction of a few $\times 10^{-3}$ in \Cthirteen.There is also a small \N14~pocket 
just atop the \Cthirteen~pocket; this region plays little role in the 
subsequent $s$-processing (\N14 is a neutron poison).

This mechanism was not found to occur very often in detailed model
calculations. Nevertheless, the fact that simple 1D convective models often show convergence problems at the bottom of the convective zone during dredge-up (\cite{FL:96}) is, we believe, an indication that the models are inadequate in this region. Semiconvection may yet be the main mechanism for activating the \Cthirteen~source.

\subsubsection{Convective Overshoot}
An obvious possible mechanism for mixing beyond a formal convective
boundary is overshooting. Note that ``overshooting'' usually refers to mixing
beyond a supposed convective-radiative boundary, which is usually determined by
the Schwarzschild or Ledoux criterion. In this case we do not mean
homogeneous mixing, or a simple extension of the convective zone into the
radiative zone. Rather, for the \Cthirteen~source to be activated we must mix 
a relatively small number of protons into the Carbon-enriched zone. 

Stimulated by the 2D hydrodynamic convective models of  \cite{Freytag:96} which showed such 
partially mixed zones, \cite{Herwig:97} introduced
an exponential decay in the convective velocity. This produces the partial mixing required 
by the \Cthirteen~source. It also, unfortunately, introduces parameters
associated with the overshooting which determine the size of the
\Cthirteen~pocket.

\subsubsection{Gravity Waves}
A recent suggestion by \cite{DT:03} is that gravity waves at the bottom of the 
convective envelope can produce partial mixing beyond the convective boundary. 
For reasonable assumptions the resultant \Cthirteen~pocket is 
about the size required 
to match observed 
abundances.

\subsubsection{Rotation}

In stellar models computed with rotation, during the AGB phase a large angular 
velocity
gradient forms, just after the occurrence of the third dredge-up, at the interface between the
faster-rotating core and the slower-rotating envelope. A zone where partial mixing of
protons and $^{12}$C is thus generated at the core/envelope interface because of shear
mixing. As with the other types of mechanism for the production of the $^{13}$C pocket, 
the proton profile is continuous in the region so that a $^{13}$C pocket is created where 
the ratio of the number of protons to $^{12}$C is below unity, and an adjacent $^{14}$N pocket is 
created where the ratio of protons to $^{12}$C is above unity. However, while with the 
other types of mechanism these two pockets keep separated during the neutron release by 
$^{13}$C($\alpha$,n)$^{16}$O in the interpulse period, in the rotating
models shear mixing persists throughout the interpulse period because the steep 
angular velocity gradient remains at the mass coordinate of the pocket formation.
Consequently, a large amount of $^{14}$N is mixed down into the $^{13}$C-rich region 
\cite{herwig:03,siess:04}. These $^{14}$N nuclei act as a strong neutron poison during the 
$s$-process because of the relatively high cross section of the $^{14}$N(n,p)$^{14}$C 
reaction. The last measurement of this reaction gave 2.04 $\pm$ 0.16 mbarn at 24.5 keV 
\cite{gledenov:95} confirming previous calculations \cite{bahcall:69} and measurements 
\cite{koehler:89}. In conclusion, the $s$-process is completely inhibited in models that 
include rotation \cite{herwig:03}. This result has been produced for a model of 3 
$M_{\odot}$ with initial surface rotation velocity of 250 km s$^{-1}$, but also 
confirmed for lower initial velocities.

Since all stars rotate, this is a major problem for the current models! One should also 
consider though that magnetic fields are generated by rotation, but have not 
been included in the computations of AGB stars so far and could represent a 
way to ``save the $s$-process''. Magnetic fields, in fact, could enhance the coupling 
of core and envelope thus decelerating the core and reducing the strength of the 
rotational shear mixing \cite{spruit:98}. The problem is open and requires further 
exploration.

If the effect of rotation could be reduced, then a spread of neutron exposures 
could be produced by different amount of $^{14}$N mixed into the $^{13}$C 
pocket as a result of a spread of rotational mixing coefficients 
\cite{herwig:03}. Then, different neutron exposures result in different final 
$s$-process abundance distributions 
\cite{siess:04}. As we describe below in more detail, several types of 
observational constraints seem to require such a spread of efficiencies to occur:

\begin{enumerate}

\item{the spectroscopic observations of the $s$-element distributions in AGB and 
post-AGB stars of different metallicities,}

\item{the spectroscopic observations of lead in stars of low metallicity,}

\item{the isotopic composition of single presolar silicon carbide grains from carbon 
stars.}     

\end{enumerate}

In summary, a possible scenario within the current models is that the spread of 
efficiencies in the $^{13}$C neutron source required by the observations is 
somewhat related to the effect of rotation and magnetic fields in stars, even 
through more defined conclusions will be set only via much future detailed work. 

\section{AGB Star $s$-Process Results}

The latest generation of $s$-process models are based on AGB stellar 
structure computed by evolutionary codes with the artificial 
introduction of a parametrized $^{13}$C pocket \cite{gallino:98,goriely:00}. The 
two phases of neutron-capture processes experienced by the 
intershell material during an interpulse-pulse 
period can be summarised as following.

\begin{description}

\item[The $^{13}$C source.] {After less than a few thousand years from the 
occurrence of proton diffusion into the intershell at the end of dredge-up, the 
$^{13}$C pocket is formed. The $^{13}$C($\alpha$,n)$^{16}$O reaction is
activated during the interpulse phase in radiative conditions at low 
temperature, $\sim 9 \times 10^{7}$ K, and all the $^{13}$C is typically consumed 
before the end of the interpulse period \cite{straniero:95}.
This is the main site for the $s$-process. The neutron flux lasts typically 10,000 yr 
and can produce very high neutron exposures, up to $\sim$ 0.5 mb$^{-1}$ in 
solar-metallicity stars, but with low neutron density values, only up to about 
$10^7$ n/cm$^{3}$ (in solar-metallicity stars).}

\item[The $^{22}$Ne source.]{At the end of the interpulse the $^{13}$C pocket is
engulfed by the following convective pulse and thus mixed with intershell
material from the previous convective pulse and the ashes from the H-burning
shell. In the convective pulse, the $^{22}$Ne($\alpha$,n)$^{25}$Mg reaction is
marginally activated at temperatures $\geq 2.5 \times 10^{8}$ K, 
and a second neutron flux occurs, whose strength depends on the temperature
at the base of the convective pulse. A large amount of $^{22}$Ne is
present in the convective pulse as a product of double $\alpha$-captures starting
on the abundant $^{14}$N from the H-burning ashes. This second neutron burst is
opposite in features to that in the $^{13}$C pocket: it occurs on a timescale of a 
few years and it produces typically very low neutron exposures, of the order of
$10^{-2}$ mb$^{-1}$ in solar metallicity stars, with a high-peaked neutron 
density, up to $10^{11}$ n/cm$^{3}$, in solar-metallicity stars.  This 
neutron burst does not contribute much to the overall production of $s$ elements, 
however it has a large effect on  the final abundances of isotopes connected to branching 
points.}

\end{description} 

After each thermal pulse the $s$-process rich material from the He intershell is 
dredged up to the envelope by the next dredge-up event. This cycle is repeated 
over all thermal pulses 
with dredge-up and the heavy element composition of the envelope throughout the AGB phase is 
changed.

\subsection{Variations with the metallicity}

As first observed by \cite{clayton:88}:
``Take the stellar structure to be independent of Z, as is the intershell $^{12}$C 
content because it is manufactured by the burning within the star. Thus the neutron 
source is in this case independent of the initial metallicity, but the major 
absorbers ($^{22}$Ne + Fe) are metallicity dependent [...] the neutron density is 
proportional to Z$^{-1}$. That is, {\it more metal-poor stars produce larger 
neutron fluences!}\,'' Current models with primary $^{13}$C as the main source
do indeed produce heavier and heavier elements at lower and lower Z, until mostly 
Pb is produced. This result is independent of the uncertainties associated with the
formation of the $^{13}$C pocket, if we assume that the mechanism that generates 
the proton diffusion in the intershell always produces a primary $^{13}$C neutron source 
in stars of different metallicities. This assumption could appear rather bold, given the 
poor knowledge of this mechanism. However, current models including this assumption 
seem to work quite well in explaining major observational features. Hence, it 
may actually turn out that this assumption {\it should} be verified by  models 
for the formation of the $^{13}$C source.

Working within this hypothesis, the following results are obtained when we keep
fixed the amount of the $^{13}$C neutron source while changing the initial
metallicity of the star. At metallicities close to solar, Sr-peak elements are
produced. At metallicities about 1/4 of solar Ba-peak elements are produced, while
at lower metallicities Pb is produced (see Figures 1 and 2 of \cite{busso:01}).

\subsection{Stellar observations at different metallicity}

The current model predictions well match the general features of the observed 
distribution of heavy elements in AGB stars of different metallicities. However, a 
spread in the efficiency of the neutron flux in the $^{13}$C pocket has to be 
introduced at each metallicity to cover the spread in the distribution shown by the 
observational data \cite{busso:01}. This spread of efficiencies in the neutron flux 
produced in the $^{13}$C pocket is represented by choices
of the $^{13}$C efficiencies ranging from the maximum neutron exposure allowed by the 
presence of the $^{14}$N poison to lower values down to zero. For example, at solar 
metallicity $\tau \leq$ 0.5 mb$^{-1}$. As discussed above, this spread 
toward neutron exposures lower than predicted can be produced by 
different efficiency in the mixing of $^{14}$N produced in the upper region of the partial 
mixing zone down to the $^{13}$C-rich region.

The Pb overabundances recently observed in stars of low metallicity represent another
important indication that the $^{13}$C neutron source is in fact of primary nature, as
the production of Pb in these stars was predicted by the current models 
\cite{gallino:98}. The distribution of the Pb overabundances and the ratio Pb/Ce show 
a spread of efficiencies at any given metallicity \cite{vaneck:03}
requiring again a spread of efficiency of the $^{13}$C neutron source at any
metallicity.

\subsection{The Galactic Chemical Evolution of heavy elements}

The consequences of applying the recent $s$-process models to the chemical 
evolution of heavy elements in the Galaxy have been presented in a series of paper 
by Travaglio, and collaborators. The main conclusions are:

\begin{itemize}

\item{The Galactic abundance of Ba-peak elements is explained 
by $s$- and $r$-processes with a contribution of 80\% to solar Ba from the $s$-process in AGB 
stars of metallicity $\sim$ 1/4 of solar \cite{travaglio:99}.}

\item{The Galactic abundance of Pb is explained by $s$- and $r$-processes with a contribution of 
90\% to the solar abundance of Pb from the $s$-process in AGB stars of metallicity $\sim$ 
1/10 of solar \cite{travaglio:01}. The classical {\it strong component} is now 
incorporated in the framework of the $s$-process in AGB stars.}

\item{Elements belonging to the Sr peak instead appear  
{\bf not} to be explained by the $s$- and $r$-processes! A special primary component 
from massive stars is needed to match observations of the light $s$-process elements at 
low metallicity. About 70\% of solar Sr comes from the $s$-process in AGB 
stars of metallicity $\sim$ 1/2 solar \cite{travaglio:04}.}

\end{itemize}

\section{Presolar SiC Grains from AGB stars}

New constraints on the $s$-process in AGB stars come from meteoritic silicon carbide 
(SiC) grains that formed in the expanding envelopes of carbon stars and contain trace 
amounts of heavy elements showing the signature of the $s$-process.
High-sensitivity laboratory measurements of the isotopic composition of trace heavy 
elements in single SiC of the size of micrometers provide constraints of  
precision never achieved before on models of the $s$-process and on 
neutron-capture cross sections. 

For example, the $^{96}$Zr/$^{94}$Zr ratio is very sensitive to the
nucleosynthesis in the convective pulse, which in turn depends on the still uncertain 
$^{22}$Ne($\alpha,n$)$^{25}$Mg reaction rate and the temperature at the base of the 
convective instability. This is because $^{96}$Zr is produced through a branching 
at $^{95}$Zr during the neutron flux of high peak neutron density 
released by the $^{22}$Ne($\alpha,n$)$^{25}$Mg reaction, while $^{94}$Zr is produced
during the main neutron flux released by the $^{13}$C($\alpha,n$)$^{16}$O 
reaction. Data from single SiC grains always show
deficits in the $^{96}$Zr/$^{94}$Zr ratio with respect to solar and point to a 
marginal activation of the $^{22}$Ne neutron source \cite{lugaro:03}, thus excluding 
intermediate-mass AGB stars as the parent stars of the grains, 
as well as the NACRE upper limit of the $^{22}$Ne($\alpha,n$)$^{25}$Mg reaction as the 
correct value for this reaction. Four presolar SiC grains show extreme deficits in 
the $^{96}$Zr/$^{94}$Zr ratios and are still unmatched by any of the current models. 

The $^{90,91,92}$Zr/$^{94}$Zr ratios on the other hand involve nuclei near closed 
neutron shells and thus depend on the main neutron exposure released by the 
$^{13}$C neutron source. The observed values for these ratios in single SiC 
grains are recovered by considering (again!) a spread of efficiencies in the 
neutron flux produced by the $^{13}$C source. 

The precision with which presolar grain data are obtained also
stimulates new measurements of neutron-capture cross sections (Koehler, P., these 
proceedings). The laboratory techniques for the analysis of presolar grains are
expanding rapidly, especially with the recent introduction of new instruments for  
material analysis targeted at the study of presolar grains: the NanoSIMS, a secondary 
ionization mass spectrometer with a primary ion beam of the size of nanometers 
\cite{hoppe:02} and the RIMS technique, resonant ionization mass
spectrometry combined to a laser-extraction technique \cite{savina:04}.
The current and future opportunities of constraining $s$-process models using 
laboratory data from the analysis of presolar grains are vast and compelling. 

\section{Conclusions}
Some details of the $s$-process in AGB stars seem now to be well understood, and
yet others remain a mystery. It is rather unsatisfactory that we still do
not know the mechanism(s) responsible for the production of the \Cthirteen~pocket.
As a consequence we do not know.how this might vary with stellar parameters, although
there are clearly indications from observations that there are variations 
from star to star. Nevertheless, with simultaneous attacks on the problem coming
from spectroscopy, grain analysis and advanced computer modelling, we may be 
optimistic that we will yet overcome our current limitations.

\vspace{0.5cm}
This work was supported by the Australian Research Council, 
and was performed under the auspices of the U.S. Department of Energy,
National Nuclear Security Administration by the University of California,
Lawrence Livermore National Laboratory under contract No. W-7405-Eng-48.

\end{document}